\documentstyle[prl,aps,epsf,multicol]{revtex}
\begin{document}
\begin{multicols}{2}

%\narrowtext

\noindent
{\large \bf Comment on ``Enhancement of the Tunneling Density of States in
Tomonaga-Luttinger Liquids''}\vspace{0.3cm}

In their recent letter \cite{OF} Oreg and Finkel'stein (OF)
calculated the electron density of states (DOS) for tunneling
into a repulsive Luttinger liquid (LL) close to the location
of an impurity. They found that the DOS $\rho(\omega)\sim\omega^{1/2g-1}$ is
not only enhanced compared to that of a pure system 
$\rho_0\sim{\omega}^{(g+1/g)/2-1}$, but also diverging at low energy if
$g>1/2$, $g$ being the standard LL parameter ($g<1$ for repulsive
interactions). Such a behavior of the DOS would have 
important experimental consequences.

In this comment we intend to show that OF's calculation 
suffers from a subtle
flaw which, being corrected, results into the DOS
\begin{equation}
\rho(\omega)\sim\omega^{1/g-1}
\label{correct}
\end{equation}
that is not only vanishing at $\omega\to 0$ but in fact
suppressed in comparison with the DOS of a pure LL.

We bosonize the
Fermi field as follows:
\begin{equation}
\psi = \sum_{a=R,L} \frac{\eta_{a}}{\sqrt{2\pi\alpha}}
{\rm e}^{\frac{i}{2}\left[\Theta/
\sqrt{g}+\epsilon_a\sqrt{g}\Phi\right]}\;,
\label{psi}
\end{equation}
where $\alpha$ is a short distance cut off, 
$R(L)$ refer to the right(left) moving component of the Fermi field, 
$\eta_R=\tau_x$, $\eta_L=-i\tau_y$, and
$\epsilon_{R(L)}=\pm $.
Here $\Phi$ and $\Theta$ are free Bose fields 
satisfying  
$\left[\Phi(x),\Theta(y)\right]=2\pi i{\rm sgn}(x-y)$.
The Pauli matrices $\tau_{x,y}$ stand to assure the correct
anticommutation relations between the right and the left
moving fields. Although these operators are often absorbed into a suitable 
definition of the Bose field, the equivalent representation
(\ref{psi}) is more convenient for our purposes.

In terms of the Bose fields, the Hamiltonian of the
system takes the form:
\begin{equation}
H=H_0+\frac{V_{bs}}{\pi\alpha}\tau_z\cos\left[\sqrt{g}
\Phi(0)\right],
\label{ham}
\end{equation}
where $H_0$ is the free field Hamiltonian and 
the second term describes the impurity backscattering.
Notice that $\tau_z$ is a conserved quantity,
so that one can set $\tau_z=1$ ($H=H_\uparrow$)
or $\tau_z=-1$ ($H=H_\downarrow$). 
%This explains why
%the operator $\tau_z$ is usually neglected in the literature.
Particular care is required when calculating correlation
functions involving 
$\tau_{x,y}$ which cause
transitions between the degenerate ground states
$\mid \uparrow(\downarrow)\rangle$ thus leading to a kind
of orthogonality catastrophe.

The retarded local electron Green function is
\begin{eqnarray*}
&&G^{(R)}(t) = -\frac{i\theta(t)}{2\pi\alpha}\left\{ 
\phantom{\left(\frac{\alpha}{\alpha + it}\right)^{\frac{1}{2g}}} \right.\\
&&\left(\frac{\alpha}{\alpha + it}\right)^{\frac{1}{2g}}
\langle \uparrow \mid \tau_+(t)\sin A(t)
\tau_-(0)\sin A(0) \mid \uparrow \rangle
\\
&+& \left. \left(\frac{\alpha}{\alpha - it}\right)^{\frac{1}{2g}}
\langle \uparrow \mid \tau_+(0)\cos A(0)
\tau_-(t)\cos A(t) \mid \uparrow \rangle
\right\},
\end{eqnarray*}
where $A=\sqrt{g}\Phi/2$ and the power law factors arise from the 
correlation of
the $\Theta$ field, which is decoupled from the impurity.
As it was correctly noticed by OF, the $\Phi$ field at the
impurity site develops a finite average value, the fluctuations around
which are massive. So the asymptotic behavior of the correlation functions
can be obtained by simply replacing $\Phi$ by its average value. 
For $V_{bs}>0$, $\langle \uparrow \mid \Phi \mid \uparrow \rangle
= \pi/\sqrt{g}$, and it is the first term of $G^{(R)}$ which 
asymptotically dominates, while for $V_{bs}<0$ it is the second term.
By neglecting phase factors, we find that in both cases
\[
t^{\frac{1}{2g}} G^{(R)}\propto 
\langle \uparrow \mid \tau_+(t)
\tau_-(0) \mid \uparrow \rangle = 
\langle \uparrow \mid {\rm e}^{iH_{\uparrow}t} 
{\rm e}^{-iH_{\downarrow}t} \mid \uparrow \rangle,
\]
which makes evident the analogy to the X-Ray edge problem. The 
above correlator can be written as 
$\langle \uparrow \mid U(t) U^\dagger(0) \mid \uparrow \rangle$, where $U$
is such that
$UH_\uparrow U^\dagger = H_\downarrow$. Since the action of $U$ is to
shift $\Phi \to \Phi + \pi/\sqrt{g}$, it is of the form
$U= \exp(i\pi J/2)$, where $J=N_R-N_L$ is the total electron current.
For $V_{bs}=0$, $J$ is conserved, but it acquires its own dynamics  
when $V_{bs}\not=0$.
In particular, $\langle J(t)J(0)\rangle = (2/\pi^2g)\ln t$ \cite{us}. 
We therefore conclude that, at large times,
$G^{(R)}(t) \sim (1/t)^{1/g}$, 
leading to the DOS Eq.(\ref{correct}). This result is in agreement with
the original analysis of Kane and Fisher\cite{K&F}.

In conclusion, our results show that the phase factors arising from
the anticommutativity of the right and left Fermi fields are not
innocuous. Ignoring them, one finds 
OF's results. Hence, the neglect of
those phase factors is the likely origin of the OF
overestimation of the DOS.

\vspace{0.3cm}

\noindent
Michele Fabrizio$^1$, and Alexander O. Gogolin$^2$\\
$^1$ International School for Advanced Studies,\\
$\phantom{^1}$ I-34014, Trieste, Italy.\\
$^2$ Department of Mathematics,\\
$\phantom{^1}$ Imperial College,\\
$\phantom{^1}$ London SW7 2BZ, United Kingdom.

\end{multicols}


\begin{references}
\bibitem{OF} Y. Oreg and A.M. Finkel'stein, Phys. Rev. Lett. {\bf 76},
4230 (1996).
\bibitem{us} See A.O. Gogolin and N. V. Prokof'ev, 
Phys. Rev. B {\bf 50}, 4921 (1994) and
M. Fabrizio and A.O. Gogolin, {\em ibid.} {\bf 51},
17 827 (1995), as well as references therein.
\bibitem{K&F} C.L. Kane and M.P.A. Fisher, Phys. Rev. Lett. {\bf 68},
1220 (1992).

\end{references}
\end{document}